# Field observation of soliton gases in the deep open ocean

Yu-Chen Lee [a,b],*, Sander Wahls [c]

[a] Delft Center for Systems and Control, Mechanical Engineering, Delft University of Technology, Mekelweg 5, Delft, 2628 CD, South Holland, The Netherlands
[b] Coastal Ocean Monitoring Center, National Cheng Kung University, University Road No. 1, Tainan, 701401, Taiwan
[c] Karlsruhe Institute of Technology, Institute of Industrial Information Technology, Hertzstraße 16, Karlsruhe, 76187, Baden-Württemberg, Germany

ABSTRACT

Soliton gases are large ensembles of random solitons with distinct characteristics arising from integrable system dynamics. They have been widely studied in theory and experiments, and were observed in natural lagoons. However, it remains an open question whether they occur naturally in the open ocean. Nonlinear ocean states containing solitons have been observed in the literature, but the dominance of solitons over other wave components required for a soliton gas has not been demonstrated. Our study provides the first field evidence of soliton gas sea states in the deep ocean, measured in Taiwan waters. The soliton energy ratio derived from the nonlinear Fourier transform (NFT) is used as a key parameter to quantify how close sea states are to soliton gases. We identify eleven measurements with extremely high soliton energy ratios. They are characterized by short-period waves with relatively small wave heights, accompanied by extreme steepness and Benjamin–Feir Index (BFI) values. These states are exceptionally rare, representing only 0.054% of our dataset. Since directional interference can artificially increase the estimated soliton energy ratio obtained from measured time series, we further apply a probabilistic directional filtering method to remove the directional interference. Three wave records from the Eluanbi station are found to retain high soliton energy ratios after the directional interference has been removed, confirming that they are indeed soliton gases.

## Introduction

Solitons are nonlinear localized waves that exhibit unique particle-like properties, retain their shape and velocity after collisions, and their amplitude directly determines their speed [1–3]. Soliton gases are solutions of integrable systems consisting of a large number of randomly distributed, nonlinearly interacting solitons [4]. The properties of a soliton gas are fundamentally different from the Gaussian linear dispersive radiation typically used to describe ocean waves [5–7]. There are kinetic equations that describe the evolution of the soliton density function characterizing the soliton gas dynamics. The first kinetic equation for a rarefied soliton gas based on the Korteweg–de Vries (KdV) equation for shallow water conditions was initially derived by Zakharov in 1971 [8]. Later, the kinetic equations for a dense soliton gas for both the KdV and nonlinear Schrödinger (NLS) equations were obtained by El and Kamchatnov in 2005 [9]. While the KdV equation applies to waves in shallow water, the NLS equation describes the complex envelope of deep water waves; the corresponding NLS solitons are thus envelope solitons [10]. The KdV and NLS equations can be solved analytically using nonlinear Fourier transforms (NFTs), which are also known as (direct) scattering transforms in the literature [2]. NFTs simplify the evolution of solutions of integrable nonlinear systems and reveal potentially hidden solitons and breathers within these solutions [11–13].

Over the last decade, soliton gases have become a fruitful research area due to their nontrivial mathematical and physical implications [14–17]. Experimental demonstrations of soliton gases were first carried out in optical fiber systems [18,19]. For water waves, an experimental realization of a soliton gas in a one-dimensional wave flume has been demonstrated for the KdV type [20], and for the NLS type [21]. In addition, field measurements of sea states exhibiting soliton turbulence have been reported under wind-wave conditions in the Currituck Sound, North Carolina [6], and in the Laguna Cáhuil, Chile [22]. These studies describe sea states dominated by KdV solitons in natural environments (both lagoons). For deep water waves, highly nonlinear NLS-type soliton dominated sea states have been observed in the Currituck Sound as well [12]. These field observations of soliton gases were all made in lagoons, and are thus not representative for the open ocean. To the best of our knowledge, no field observations of soliton gases in the open ocean have been reported so far.

* Corresponding author at: Delft Center for Systems and Control, Mechanical Engineering, Delft University of Technology, Mekelweg 5, Delft, 2628 CD, South Holland, The Netherlands.
*E-mail address:* yclee.delft@gmail.com (Y.-C. Lee).

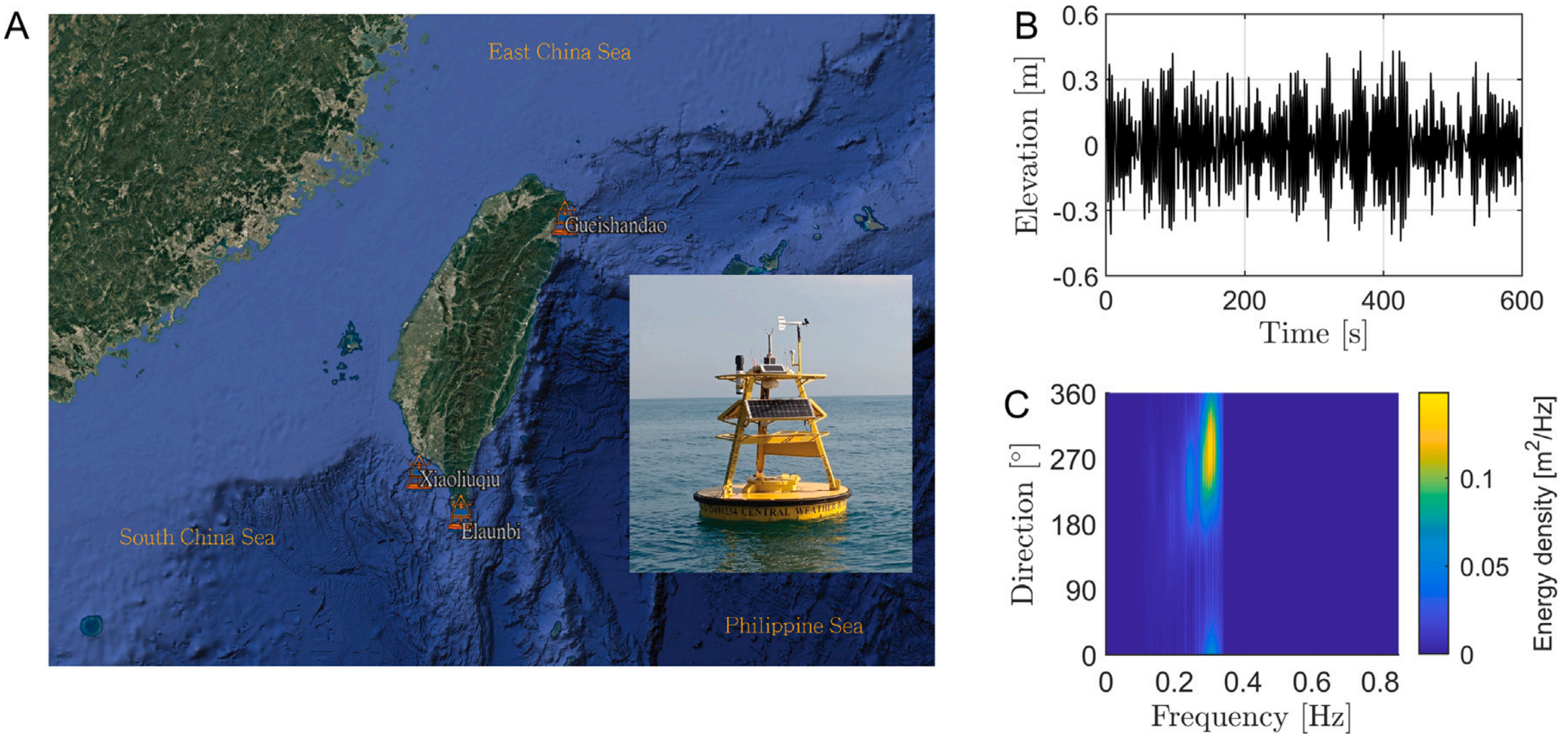

**Fig. 1.** Data collection from buoy measurements in Taiwanese waters. (A) Locations of three buoy stations, Eluanbi (south), Gueishandao (northeast) and Xiaoliuqiu (southwest). (B) Example of a soliton gas (surface elevation time series) recorded at the Eluanbi station on 2019/05/16 at 14:00. (C) Corresponding directional wave spectrum.

**Table 1**
Overview of wave measurement buoy stations and basic parameters in Taiwan waters.

| Station | Latitude | Longitude | Depth (m) | Year(s) | Data No. | $kh$ | $H_s$ (m) | $T_p$ (s) | $\sigma_\theta$ (°) |
|---|---|---|---|---|---|---|---|---|---|
| Eluanbi | 21°55′06″ N | 120°48′57″ E | 40 | 2017, 2019 | 5 977 | 1.36–18.61 | 0.17–5.32 | 3.31–10.71 | 53.86–80.33 |
| Gueishandao | 24°50′55″ N | 121°55′53″ E | 38 | 2018–2019 | 5 285 | 1.36–15.82 | 0.19–6.72 | 3.77–10.53 | 57.89–80.76 |
| Xiaoliuqiu | 22°18′46″ N | 120°21′42″ E | 82 | 2017–2019 | 9 261 | 1.36–38.15 | 0.02–6.35 | 3.77–15.38 | 51.14–80.17 |

In this study, we aim to identify soliton gas sea states under intermediate and deep water conditions in buoy measurements collected in Taiwanese waters. The buoys are operated by the Coastal Monitoring Center at National Cheng Kung University, Tainan, Taiwan. A key challenge in identifying soliton gases in the open ocean from time series is the presence of directional interference, which is not captured by the NLS equation underlying the NFT considered in this paper. We address this issue using a specialized directional filtering technique applied to simulated wave fields, which are based on directional power spectra provided by the buoys.

This paper is organized as follows. Section “Data acquisition” describes the data acquisition process. Section “Method” introduces the theoretical framework, including the nonlinear Schrödinger equation, the nonlinear Fourier transform, and the calculation of soliton parameters. Section “Results” presents the main results on the identification and statistical analysis of soliton gas sea states, including a discussion on directional interference. Finally, Section “Conclusion and Outlook” summarizes the findings and offers concluding remarks.

## Data acquisition

The surface wave data used in this study were measured by buoys operated by the Coastal Ocean Monitoring Center (COMC) at National Cheng Kung University (NCKU), Taiwan [23–25]. The discus buoy is equipped with an accelerometer-tiltcompass (ATC) sensor that can measure six degrees of freedom. Measured buoy accelerations in six degrees of freedom are used to calculate directional wave spectra by applying a Fourier transform to the cross-correlation of the motion signals [24,26]. Three buoy stations named Eluanbi, Gueishandao and Xiaoliuqiu are located in the south, northeast and southwest of Taiwan, respectively. The water depths at the three stations are 40 m, 38 m and 82 m, corresponding to intermediate and deep water conditions. The locations of the buoy stations and the buoy type are shown in Fig. 1A. An example of a time series after data quality control is presented in Fig. 1B, along with its corresponding directional wave spectrum in Fig. 1C. Table 1 presents buoy information and also includes the measured ranges of relative water depth, significant wave height, peak period, and directional spreading for each station. These data span conditions ranging from normal sea states to extreme events driven by local winds and large-scale Pacific systems, including monsoons and typhoons, thereby capturing a wide diversity of wave fields in the Pacific region. The data used in this study were collected between 2017 and 2019. The measured accelerations from six degrees of freedom were analyzed using preprocessing methods involving noise filtering to obtain time series of wave elevation, following Appendix A in Ref. [25]. The data were recorded hourly, with each record having a duration of 10 min. An example, measured at the Xiaoliuqiu buoy on 2018/04/22 at 16:00, is shown in Fig. 1B of the main manuscript. The sampling rate of the time series is 1.7 Hz. The directional wave spectrum from the ATC data was obtained following Appendix B in Ref. [25].

The nonlinear Schrödinger (NLS) equation assumes deep water conditions with narrow-banded and unidirectional wave fields, which do not always correspond to real ocean conditions. Therefore, the data are specifically selected to better satisfy NLS assumptions. First, the applicability of the NLS equation requires $kh > 1.363$ for deep water conditions, where $k$ is the peak wave number and $h$ is the water depth. (This condition marks the focusing regime of the NLS equation. Only in the focusing regime, do conventional “bright” soliton solutions exist.) We select data based on this threshold to ensure appropriate depth conditions. Second, to reduce spectral broadness in both frequency and directional domains, we retain only single-peak spectra and filter out data with double or multi-peak spectra. The coexistence of local wind seas and distant swells often results in double- or multi-peaked wave spectra [27]. This phenomenon is quite dominant in the real ocean but still depends on local climate and environmental conditions. Significant peaks in the one-dimensional wave spectra were identified using the method described in Ref. [28]. This approach limits the dataset to single wave systems, thereby minimizing the effects of spectral bandwidth that can violate the narrow-band assumption fundamental to NLS equation validity. Based on the above selection criteria, we obtained 5977, 5285 and 9261 deep water unimodal spectra for Eluanbi, Gueishandao and Xiaoliuqiu stations, respectively. These represent 47.0%, 34.5% and 42.7% of the total datasets for each station.

## Method

### *Nonlinear Schrödinger equation*

Different nonlinear Fourier transforms (NFTs) exist to deal with different types of integrable dynamics. In this paper, we consider the NFT for the NLS equation [2,29,30]. The NLS equation describes the unidirectional propagation of complex wave envelopes in deep water under the assumptions of narrow bandwidth and weak nonlinearity. It is typically expressed in spatial form, but since this study uses field measurement data in time series, the temporal NLS formulation is required. The temporal NLS equation can be expressed as [11,12,31]:

$$i\left[C_g^{-1}A_t + A_x\right] + \mu C_g^{-3}A_{tt} + \nu C_g^{-1}|A|^2 A = 0. \tag{1}$$

where $A(x,t)$ is the complex envelope, $C_g$ is the group velocity, $\mu$ is the second-order dispersion coefficient, $\nu$ is the nonlinear coefficient, $x$ is the spatial coordinate, and $t$ is time. We used the coefficients for finite depth, which are quite complicated and can be found, e.g., in [12,32]. The complex envelope is obtained from the surface elevation by taking the Hilbert transform and then removing the carrier [11,31]. Conversely, the surface elevation can be recovered from the complex envelope by adding the carrier back in and taking the real part of the result. The NFT assumes that the data are governed by the following normalized form of the NLS equation:

$$iu_X + u_{TT} + 2|u|^2 u = 0, \tag{2}$$

where $u(X,T)$ is the normalized complex envelope, $X$ is the normalized spatial coordinate, and $T$ is the normalized temporal coordinate. The change of coordinates is

$$\begin{aligned} X &= \frac{\mu}{C_g^3}x, \quad T = t - \frac{x}{C_g}, \\ u(X,T) &= \rho A(x,t), \quad \rho := \sqrt{\frac{C_g^2 \nu}{2\mu}}. \end{aligned} \tag{3}$$

where $\rho$ is the nonlinear parameter. This change of coordinates is applied before the data are passed to the NFT.

### *Nonlinear Fourier transform*

Since the direct spatial evolution of the complex envelope governed by the NLS equation can be complicated, the nonlinear Fourier transform (NFT) provides a major advantage because the evolution of the nonlinear Fourier components follows simple analytic formulas. By applying the inverse nonlinear Fourier transform (iNFT) to the propagated spectrum, one can efficiently obtain the solution at different spatial points [2,33]. In this study, we used the NFT for vanishing signals since solitons are vanishing waveforms. They have a natural representation in the nonlinear spectrum of the vanishing NFT. The NFT for periodic signals, in contrast, can only approximate solitons through cnoidal waves with very high modulus [34]. It has been shown that both approaches in general return the same soliton content [34,35], but the determination of the soliton amplitudes with the periodic NFT requires the computation of a heuristic "reference level", which can be difficult to use in practice [36,37]. We also remark that time series measured in the real ocean are generally not strictly periodic. We utilized the open-source software library 'Fast Nonlinear Fourier Transforms' (FNFT) for the numerical computations [31,38]. This library provides efficient numerical algorithms for the computation of various types of NFTs. We now outline the computation of the NFT for the NLS equation [29,30].

The scattering problem for the NLS equation is

$$\Psi_T = Q(\zeta)\Psi, \quad Q = \begin{pmatrix} -i\zeta & u(X_0,T) \\ -u(X_0,T)^* & i\zeta \end{pmatrix}, \tag{4}$$

where $\zeta$ is the complex spectral parameter, $^*$ represents complex conjugation, and $X_0$ is an arbitrary spatial reference point for the time series. Two pairs of linearly independent solutions called Jost functions, $[\phi(T,\zeta) \quad \bar{\phi}(T,\zeta)]$ and $[\psi(T,\zeta) \quad \bar{\psi}(T,\zeta)]$, are found by solving (4) with

$$\begin{aligned} &\phi(T,\zeta) \to \begin{bmatrix} e^{-i\zeta T} \\ 0 \end{bmatrix}, \ \bar{\phi}(T,\zeta) \to \begin{bmatrix} 0 \\ -e^{i\zeta T} \end{bmatrix} \quad \text{as} \quad T \to -\infty, \\ &\psi(T,\zeta) \to \begin{bmatrix} 0 \\ e^{i\zeta T} \end{bmatrix}, \ \bar{\psi}(T,\zeta) \to \begin{bmatrix} e^{-i\zeta T} \\ 0 \end{bmatrix} \quad \text{as} \quad T \to \infty. \end{aligned} \tag{5}$$

The pairs of Jost functions are related through the scattering coefficients $a(\zeta)$, $b(\zeta)$, $\bar{a}(\zeta)$, $\bar{b}(\zeta)$:

$$\begin{aligned} \phi(T,\zeta) &= a(\zeta)\bar{\psi}(T,\zeta) + b(\zeta)\psi(T,\zeta), \\ \bar{\phi}(T,\zeta) &= \bar{b}(\zeta)\bar{\psi}(T,\zeta) + \bar{a}(\zeta)\psi(T,\zeta). \end{aligned} \tag{6}$$

The scattering coefficients $a(\zeta)$ and $b(\zeta)$ are the basis for forming the nonlinear spectrum, which has a continuous part and a discrete part. The continuous spectrum is described by the *reflection coefficient*

$$r(\zeta) = b(\zeta)/a(\zeta), \quad \zeta \in \mathbb{R}. \tag{7}$$

It represents the radiation components in the time series. The discrete spectrum contains the discrete eigenvalues $\{\zeta_n = \xi_n + i\eta_n\}$, $n = 1, \ldots, N$, where $N$ is the total number of solitons in the signal. They are the zeros of the scattering coefficient $a(\zeta)$ with positive imaginary part. The eigenvalues are complemented by the residues

$$r_n = b(\zeta_n)/a'(\zeta_n), \quad \text{where} \quad a' = da/d\zeta. \tag{8}$$

Each eigenvalue in the discrete spectrum corresponds to one solitonic component in the time series.

The inverse NFT is, in contrast to the linear Fourier transform, quite different from the forward NFT. There are several ways to perform the inversion. In our case, where only the discrete spectrum containing the solitons is being inverted, FNFT employs the so-called Darboux transform to recover the time domain signal [39].

### *Soliton amplitudes and velocities*

We now describe how the physical soliton amplitudes and velocities can be obtained from the discrete part of the NFT, which has been computed from normalized data. The magnitude of an envelope soliton arising from a single complex eigenvalue $\zeta_n = \xi_n + i\eta_n$ is [2]

$$|u(X,T)| = 2\eta_n \,\mathrm{sech}\left[2\eta_n(T + 4\xi_n X)\right]. \tag{9}$$

Reverting the data normalization discussed earlier, we bring this soliton into physical dimensions:

$$\begin{aligned} |A(x,t)| &= \left|\frac{u(X,T)}{\rho}\right| = \left|\frac{1}{\rho}u\left(\frac{\mu}{C_g^3}x,\ t - \frac{x}{C_g}\right)\right| \\ &= \frac{2\eta_n}{\rho}\,\mathrm{sech}\left[2\eta_n\left(t - \left(\frac{1}{C_g} - 4\xi_n\frac{\mu}{C_g^3}\right)x\right)\right] \\ &= \frac{2\eta_n}{\rho}\,\mathrm{sech}\left[2\eta_n\frac{-1}{C_{\mathrm{sol}}}\left(x - C_{\mathrm{sol}}\,t\right)\right]. \end{aligned} \tag{10}$$

where

$$C_{\mathrm{sol}} = \frac{C_g}{1 - 4\xi_n\mu/C_g^2} \approx C_g + \frac{4\xi_n\mu}{C_g}. \tag{11}$$

is the soliton velocity and $2\eta_n/\rho$ is the soliton amplitude, both in physical dimensions. The approximation holds under the condition $4\xi_n\mu/C_g^2 \ll 1$.

*Soliton energy ratio calculation*

To detect envelope solitons in the measured time series, the complex envelope is first computed and normalized. In general, NFTs are a class of transforms that decompose time (or space) series into physically meaningful components w.r.t. integrable dynamics [30]. They play a key role in the inverse scattering method to solve integrable partial differential equations and are also known as forward scattering transforms. In light of their unique capability to uncover hidden solitons, NFTs have been applied successfully to many real-world systems, even if the underlying physical models are not fully integrable [11,40–44]. In particular, they are widely used to study soliton gases. With the NFT for the NLS employed here, solitons correspond to discrete spectral components, while dispersive radiation waves are associated with a continuous spectrum. The soliton energy ratio is found from the NFT and lets us quantify the soliton content in time series. It is based on the nonlinear Parseval formula [2]

$$E_{\text{total}} = \int_{-\infty}^{\infty} |u(X_0, T)|^2 dT = E_{\text{sol}} + E_{\text{rad}}, \tag{12}$$

where we set $u(X_0, T) = 0$ for times outside the measurement interval. $E_{\text{sol}}$ is the energy of the solitons, and $E_{\text{rad}}$ is the energy of the radiation waves. The energy of the radiation waves can be calculated using the reflection coefficient:

$$E_{\text{rad}} = \frac{1}{\pi} \int_{-\infty}^{\infty} \log(1 + |r(\zeta)|^2) d\zeta \tag{13}$$

is computed from the continuous part of the NFT and represents the energy in the radiation components, and

$$E_{\text{sol}} = 4 \sum_{n=1}^{N} \eta_n^2 \tag{14}$$

is computed from the discrete part of the NFT and represents the energy in the solitonic components. The soliton energy ratio can be computed as:

$$E_{\text{sol.ratio}} = \frac{E_{\text{sol}}}{E_{\text{total}}} = \frac{E_{\text{total}} - E_{\text{rad}}}{E_{\text{total}}} \in [0, 1]. \tag{15}$$

This serves as a quantitative measure of the degree of nonlinearity in the sea state. It is zero in the complete absence of solitons, and one for a pure soliton gas. Soliton gases are thus indicated by high soliton energy ratios.

*Statistical wave parameters for sea state characterization*

We discuss the wave parameters used to statistically quantify sea state properties. To analyze surface wave elevations, we use the downward zero-crossing method [45]. This involves identifying individual waves by finding points where the surface falls below its average level. From the waves in the record, we calculate the maximum wave height $H_{\text{max}}$, which is the largest individual height. The significant wave height $H_{1/3}$ represents the average height of the highest one-third of all waves within the record found with the downward zero-crossing method. It is a statistical parameter used to characterize the wave conditions of a sea state. The abnormality index (AI) quantifies the maximum wave height compared to the significant wave height [46, 47]:

$$\text{AI} = \frac{H_{\text{max}}}{H_{1/3}}. \tag{16}$$

Waves with an abnormality index AI $\geq 2$ are classified as rogue waves.

The directional spreading quantifies the angular distribution of wave energy around the peak direction and is calculated using directional moments [48,49]:

$$\sigma_\theta = \left[ 2 \left\{ 1 - \left( \frac{p^2 + q^2}{m_0^2} \right)^{\frac{1}{2}} \right\} \right]^{\frac{1}{2}}, \tag{17}$$

where $p = \int_0^{2\pi} \int_0^{\infty} \cos(\theta) S(\omega, \theta)\, d\omega\, d\theta$, $q = \int_0^{2\pi} \int_0^{\infty} \sin(\theta) S(\omega, \theta)\, d\omega\, d\theta$, $m_0 = \int_0^{2\pi} \int_0^{\infty} S(\omega, \theta)\, d\omega\, d\theta$ is the zeroth spectral moment, and $S(\omega, \theta)$ is the directional spectrum, representing the distribution of surface wave energy across both angular frequency $\omega$ and direction $\theta$.

Kurtosis quantifies the fourth standardized moment of the surface elevation time series [50]. Higher kurtosis values indicate an increased probability of encountering extreme wave events. Skewness, the third standardized moment of the surface elevation time series, measures the asymmetry of the surface elevation distribution, indicating the vertical asymmetry between wave crests and troughs [50]:

$$\text{Kurtosis} = \frac{\overline{\eta^4}}{\sigma^4}, \tag{18}$$

$$\text{Skewness} = \frac{\overline{\eta^3}}{\sigma^3}, \tag{19}$$

where $\eta$ is the surface elevation, and $\sigma$ is the standard deviation. Wave steepness characterizes the ratio of significant wave height to wavelength and is commonly expressed as [51]

$$\text{Steepness} = \frac{2\pi H_s}{g T_p^2}, \tag{20}$$

where $g$ is gravitational acceleration. Here, $H_s$ denotes four times the standard deviation of the surface elevation. It is another commonly used parameter for the significant wave height. Under certain statistical assumptions, it coincides exactly with $H_{1/3}$. The peak period $T_p$ is defined as the wave period corresponding to the maximum spectral energy density. The Benjamin-Feir Index (BFI) measures the potential for modulational instability and nonlinear wave interactions [44,50,52]:

$$\text{BFI} = \frac{\sqrt{2}\varepsilon}{\delta_\omega}, \tag{21}$$

where $\varepsilon$ is the wave steepness, $\delta_\omega$ is the spectral bandwidth calculated from Goda's peakedness parameter [53]. The BFI can be interpreted as a ratio between nonlinear focusing and linear dispersive spreading. A large BFI means that nonlinear effects are strong compared to spectral spreading, so the sea state is more susceptible to modulational instability and soliton-like dynamics.

## Results

*Identifying soliton gas sea states*

Using field measurement data from three buoy stations, we applied the NFT and calculated the soliton energy ratio for records that satisfied both unimodal spectrum and deep water criteria. Selecting a unimodal spectrum restricts the wave field to the influence of a single wind system, thereby avoiding crossing seas and broadened spectral bandwidth. We initially looked for seas that exhibit a very high soliton energy ratio of at least 0.9. Based on this criterion, we identified 3, 2 and 6 events among the 5977, 5285 and 9261 qualified records from Eluanbi, Gueishandao and Xiaoliuqiu stations, respectively. These correspond to occurrence rates of 0.050%, 0.038% and 0.065% under unimodal spectrum and intermediate and deep water conditions. Although some records were in intermediate depth, the soliton gases identified in this study fully satisfy the deep water conditions. In total, 11 time series with very high soliton energy ratios were identified from 20,523 records from three buoy stations, which yields an overall occurrence rate of 0.054%. This indicates that sea states with very high soliton energy ratios are extremely rare events, at least for the high soliton energy ratio cutoff considered here. To our knowledge, such sea states have not been previously observed in the literature. We note that buoy measurements tend to underestimate extreme wave heights and suppress higher wave crests compared to radar observations [54,55]. This study, which is based on buoy data, therefore likely underestimates the occurrence of soliton gases. The application of the same method to radar measurements should reveal a greater prevalence.

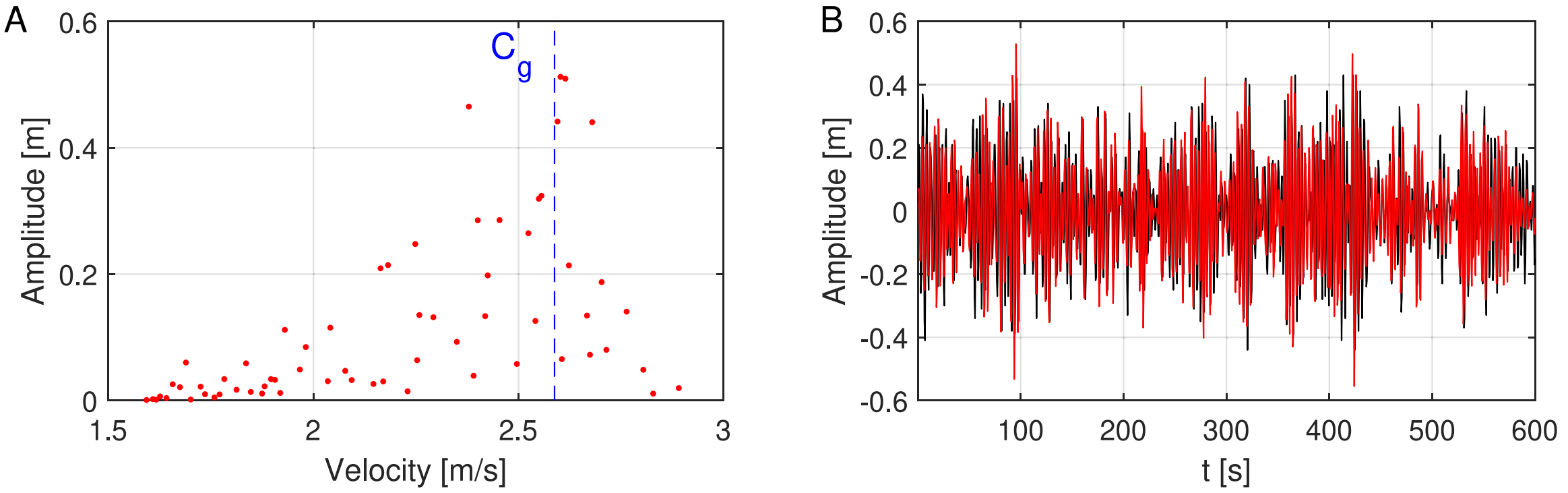

**Fig. 2.** NFT analysis of a sea state with high soliton energy ratio. (A) Soliton spectrum. (B) Corresponding initial surface wave elevation (black) compared with soliton time series (red) reconstructed from the soliton spectrum.

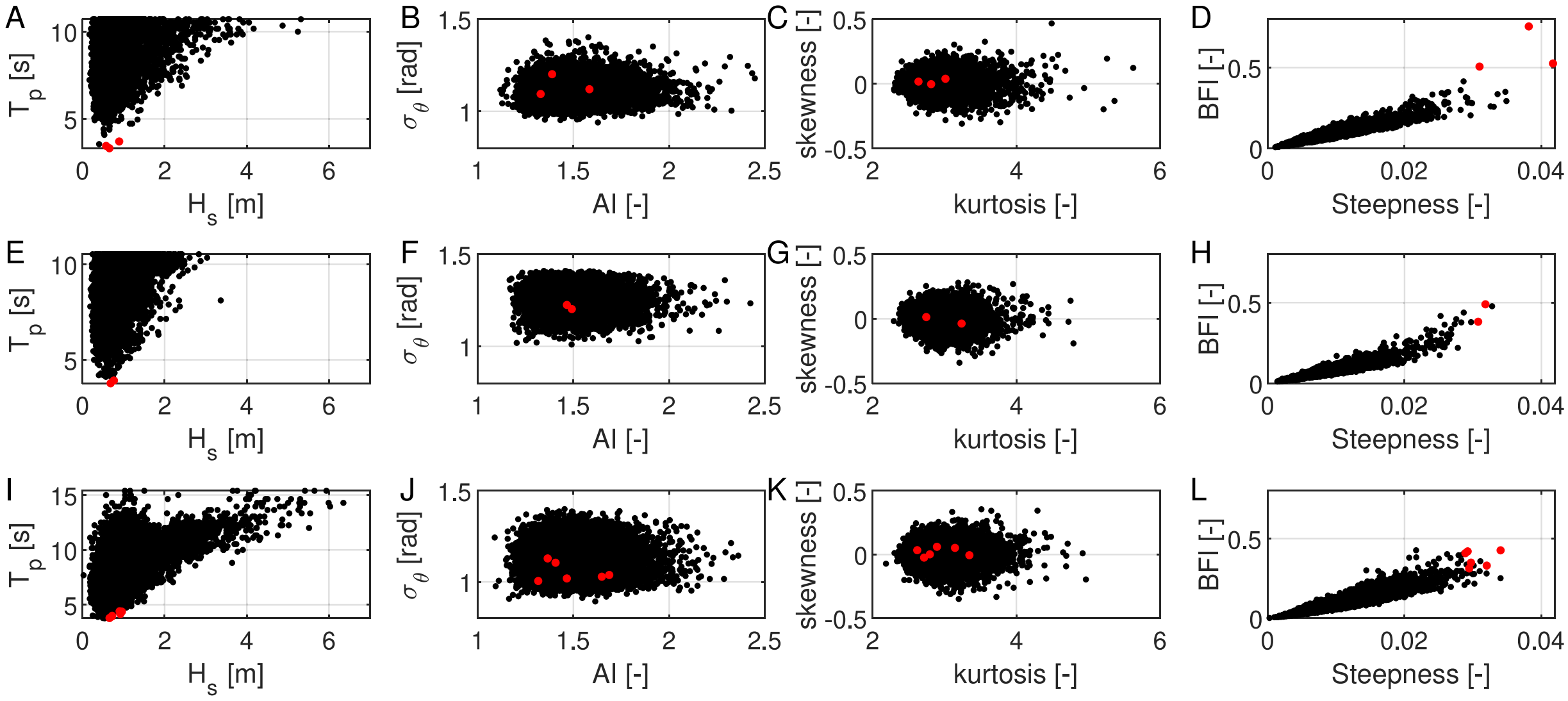

**Fig. 3.** Statistical properties of sea states with high soliton energy ratios of at least 0.9 (red) compared with all data (black) at Eluanbi (A–D), Gueishandao (E–H), and Xiaoliuqiu (I–L). (A), (E), (I): scatter plots of significant wave height $H_s$ versus peak period $T_p$; (B), (F), (J): abnormality index ($AI$) versus directional spreading $\sigma_\theta$; (C), (G), (K): kurtosis versus skewness; (D), (H), (L): wave steepness versus Benjamin–Feir Index (BFI).

Fig. 1B shows one of the identified time series with very high soliton energy ratio, measured on 2019/05/16 at 14:00 at the Eluanbi station; the corresponding directional spectrum is shown in Fig. 1C. The complex envelope of the surface wave elevation time series was obtained using the Hilbert transform [56], which then served as input for the NLS-NFT after the carrier frequency had been removed. The resulting soliton spectrum is shown in Fig. 2A. Each red dot corresponds to a point in the discrete part of the NFT and indicates a soliton. A total of 64 solitons were detected. These axes were transformed back to the physical domain. The largest soliton amplitude is 0.51 m. The soliton energy ratio is 0.96. This soliton spectrum is typical for seas with high soliton energy ratio. To illustrate the dominance of solitons in the time domain, we applied the inverse NFT to the discrete spectrum only. By not considering the continuous spectrum in the inverse NFT step, we filter the signal nonlinearly and obtain the complex envelope of the underlying solitonic part. Using the same carrier frequency as the initial time series, the surface elevation of the soliton components was reconstructed (red line in Fig. 2B). The reconstructed soliton time series (red line) closely matches the original measurement (black line), confirming the very high soliton energy ratio. The residual radiation wave components may either amplify or attenuate the soliton part.

*Statistical properties of soliton gas sea states*

To further explore the characteristics of the 11 sea states with high soliton energy ratios (above 0.9), we analyze the statistical properties of various wave parameters including significant wave height, peak period, directional spreading, abnormality index (AI), kurtosis, skewness, steepness, and the Benjamin–Feir Index (BFI). The statistical distributions of these parameters for the eleven sea states are compared with the overall dataset, as shown in Fig. 3, where (A)–(D), (E)–(H), and (I)–(L) correspond to data from the Eluanbi, Gueishandao and Xiaoliuqiu stations, respectively. In the figure, red dots represent very high soliton energy ratios, while black dots denote the entire dataset. We first focus on the scatter plots of significant wave height versus peak period (Fig. 3A, E, I), which reveal that these seas have relatively smaller wave heights and shorter peak periods. This observation deviates from the concept that individual solitons or breathers can exhibit their characteristic shapes in time series, forming large-amplitude rogue waves [44,57]. However, our data share similar characteristics of small wave heights and short peak periods with highly nonlinear breather turbulence wind waves measured in Currituck Sound, North Carolina [12]. We then examine the scatter plots of AI versus directional spreading (Fig. 3B, F, J). It can be observed that the seas with high soliton energy ratios do not exhibit large AI values, indicating that rogue waves are absent under these conditions. Moreover, these seas are not correlated with directional spreading. Subsequently, we analyze kurtosis versus skewness (Fig. 3C, G, K). The results show that seas with high soliton energy ratios do not display large kurtosis. However, they exhibit skewness values close to zero, indicating that the time series are nearly symmetric. The most interesting results are the scatter plots of steepness versus BFI (Fig. 3D, H, L), which reveal that sea states with high soliton energy ratios exhibit exceptionally high wave steepness and BFI values. Specifically, all of these events have steepness exceeding 0.029 and BFI values above 0.31. As expected, the results confirm that seas with high soliton energy ratios are characterized by high wave steepness and BFI, a finding consistent with their strong nonlinearity. We additionally note that the wave ages of highly nonlinear soliton dominated seas range from 0.30–0.45 at Eluanbi, 0.63–0.65 at Gueishandao, and 0.60–0.90 at Xiaoliuqiu. These values all correspond to young wind seas, indicating that they represent wind-generated waves consistent with the findings of Osborne et al. (2019) [12].

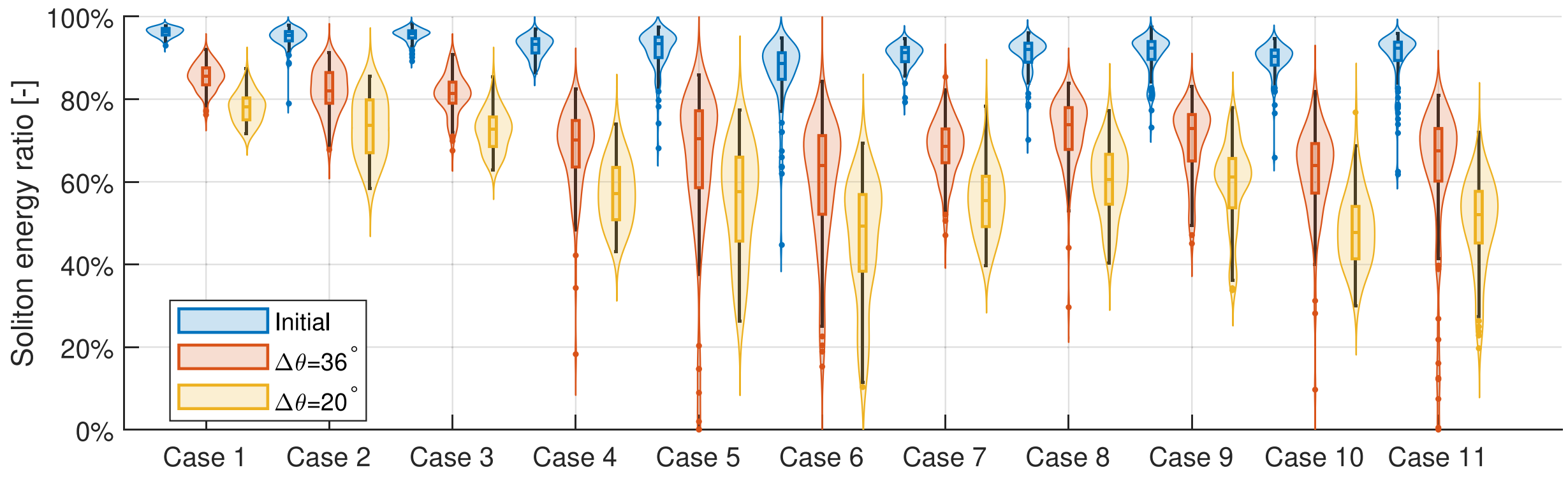

**Fig. 4.** Investigation of reduced directional effects on field-measured soliton gases: distribution of soliton energy ratios for the 11 soliton gas cases after directional filtering with 100 sets of random phases per case. Violin plots show the analyzed soliton energy ratio for: the initial wave field (blue), retention angle $\Delta\theta = 36°$ (orange), and retention angle $\Delta\theta = 20°$ (yellow).

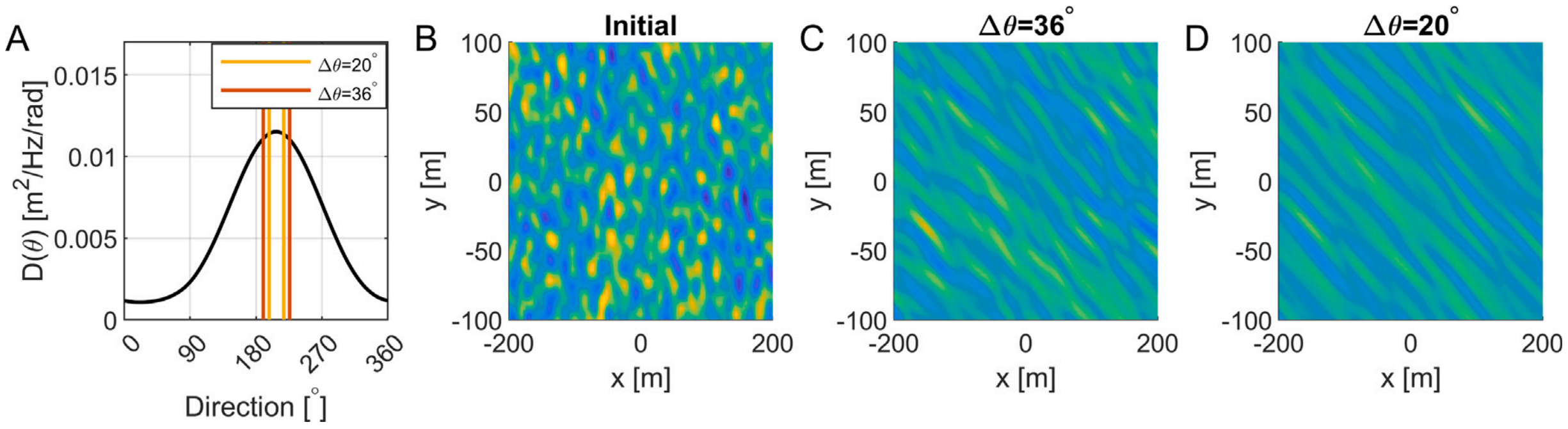

**Fig. 5.** Investigation of reduced directional effects on field-measured soliton gases (A) Directional spectrum with retention angles $\Delta\theta = 36°$ (orange lines) and $\Delta\theta = 20°$ (yellow lines) with energy outside the angles truncated. (B) Simulated wave field from random phases with the full spectrum. (C–D) Reconstructed wave fields with the same phases but truncated spectra at $\Delta\theta = 36°$ and $\Delta\theta = 20°$, respectively.

*Discussion on directional interference in soliton gas seas*

Integrable models like the KdV or NLS equation describe real-world ocean waves only approximately. Directional interference in particular can suppress the dynamics of NLS soliton gases by disrupting nonlinear focusing mechanisms and redistributing energy across different propagation directions. Such suppressed soliton gases represent states in which initially detected solitons are subsequently weakened by directional spreading. In simulations of directional JONSWAP wave fields, the soliton energy ratio computed directly from time series was found to typically overestimate the soliton content in the main propagation direction due to directional interference, although the effect was less pronounced for strongly nonlinear sea states [58]. To ensure that the soliton energy ratios computed from the time series in our dataset are not overestimated, we reconstruct the part of the wave fields propagating in the main direction under the linear assumption. Since only the magnitudes of the directional spectrum are available in our data, we choose the phases uniformly at random. To ensure representativeness, many different sets of random phases are considered. The directional magnitude spectrum of each simulated time series coincides with that of the original time series in the main propagation direction, but the directional interference has been removed. By computing the soliton energy ratios for different realizations of the random phases, we obtain a probability distribution of the possible soliton energy ratios for the true time series without directional interference.

In our data acquisition, we focus on cases with a unimodal spectrum to minimize spectral width and directional effects. However, for field measurement data, the directional spreading can still be substantial. Directional interference may lead to an overestimation of the soliton energy ratio, as wave components from multiple directions contribute spurious energy to the single-point measurement, which does not reflect the true soliton content in the primary propagation direction. To address this, we further examine how the soliton energy ratio changes when only waves propagating in the main direction are considered. Here, we apply directional truncation with two retention angles, $\Delta\theta = 36°$ and $\Delta\theta = 20°$, whereby energy outside the retention angle of the main spectral peak is excluded.

To reconstruct surface wave fields from a given directional wave spectrum, the surface is modeled as a linear superposition of small-amplitude oscillations with random phases [59–62]. Specifically, it is represented as a superposition of many simple cosine waves, each with varying amplitude, frequency, phase, and propagation direction. The sea surface elevation $\eta(x,y,t)$ at a spatial location $(x,y)$ and time $t$ is then given by

$$\eta(x,y,t) = \sum_{i=1}^{N_\omega}\sum_{j=1}^{N_\theta}\sqrt{2S(\omega_i,\theta_j)\Delta\omega\Delta\theta} \cos\left(k_i(x\cos\theta_j + y\sin\theta_j) - \omega_i t + \phi_{ij}\right), \quad (22)$$

where $N_\omega$ and $N_\theta$ are the total number of discrete frequency and direction bins, respectively. The term $\Delta\omega$ is the angular frequency increment, and $\Delta\theta$ is the directional increment. The directional wave density spectrum $S(\omega_i,\theta_j)$ is derived from buoy measurements for each time series. The wave number $k_i$ corresponding to frequency $\omega_i$ is calculated using the deep water dispersion relation. Since the phases $\phi_{ij}$ have not been measured, they are chosen randomly from a uniform distribution in the interval $[0,2\pi]$. To capture the influence of the random phases, 100 sets of random phases are considered for each time series.

For the spatial domain, we set $x \in [-200,200]$ m and $y \in [-100,100]$ m, with a grid resolution of $N_x = 101$ and $N_y = 51$. The time is initialized at $t = 0$, with a 0.59 s interval, for a total duration of 600 s. Subsequently, we generate wave fields constrained to specific retention angles, where the energy of the directional spectrum outside the peak retention range is removed. To ensure that only the components in the main direction are extracted, the same random phases as those of the initial wave field are retained. Under the assumption of linear superposition, this approach enables the successful reconstruction of wave fields propagating exclusively in the main direction. The other directions thus contribute additive perturbations to the main direction in our framework, which is a standard model in the study of transversal perturbations of line solitons [63,64]. However, note that the reconstructed main direction may still contain features that arose from nonlinear interactions with the rest of the wave field. Moreover, the use of wave numbers obtained from the deep water dispersion relation means that nonlinear dispersion effects and bound-wave contributions are not explicitly represented.

As an illustration, we present an example based on one of our seas with high soliton energy ratios in Fig. 5. In this case, the directional spreading is $\sigma_\theta = 57.58°$, and the corresponding directional spectrum is shown in Fig. 5A. The retention angles of $\Delta\theta = 36°$ and 20° are indicated by the orange and yellow lines, respectively. Under the linear superposition of waves with random phases, the simulated wave field corresponding to the initial spectrum is shown in Fig. 5B. By removing the energy outside the retention angles in the directional sidebands while preserving the original phases, we reconstructed the wave fields, as shown in Fig. 5C for $\Delta\theta = 36°$ and Fig. 5D for $\Delta\theta = 20°$. It can be seen that after the removal of energy in the directional sideband, the initial short-crested wave field changes into long-crested wave fields. A virtual wave gauge was placed at the center of the domain ($x = 0$, $y = 0$) to extract simulated time series for nonlinear Fourier analysis.

We applied the above method to the 11 states with high soliton energy ratios identified before to examine how the removal of directional interference in the sidebands affects the soliton energy ratio. Slunyaev (2024) recently observed that solitons can persist in broadbanded and moderately directional sea states ($\Delta\theta \le 30°$) in deep water, where they remained stable for several tens of wave periods [65]. With less directional spreading, their lifetime can be even longer. In light of these results, wave energy outside the principal wave directions was removed with retention angles of $\Delta\theta = 36°$ and $\Delta\theta = 20°$; only energy within these ranges was preserved. Since the directional phases were unknown, we generated 100 random phase realizations for each measured directional (power) spectrum to construct simulated wave fields. The resulting soliton energy ratios for all cases under different retention angles $\Delta\theta$ are shown in Fig. 4. They are depicted as violin plots that indicate the distribution of the soliton energy ratios resulting from considering 100 sets of random phases. The "initial" cases correspond to the initial, unfiltered directional spectrum equipped with random phases. The cases 1–3 were measured at the Eluanbi station, the cases 4–5 at the Gueishandao station, and the cases 6–11 at the Xiaoliuqiu station. For simulations using the full directional spectrum (blue boxes), most reconstructed time series have mean soliton energy ratios exceeding 0.9. As expected, the mean soliton energy ratio decreases after directional truncation. This is because removing energy from the system leads to a reduction in its nonlinearity. A larger retention angle ($\Delta\theta = 36°$) leads to a moderate reduction in the mean soliton energy ratio (orange boxes), while a smaller retention angle ($\Delta\theta = 20°$) results in a stronger reduction in the mean soliton energy ratio (yellow boxes). Averaged over all simulations, the mean soliton energy ratio is 0.92 for the full spectrum (blue boxes), decreases to 0.70 at $\Delta\theta = 36°$ (orange boxes), and further decreases to 0.60 at $\Delta\theta = 20°$ (yellow boxes). For the cases 1 to 3 (Eluanbi station), the distributions stay completely above the threshold of 0.5, above which solitons are dominating. Furthermore, as shown in Table 2, these three cases also satisfy the other assumptions of the NLS equation sufficiently well. *We can thus consider them soliton gases even with directional interference taken into account.* For the cases 4 to 11 (Gueishandao and Xiaoliuqiu), the distributions in Fig. 4 reach below 0.5. Thus, with directional interference taken into account, we can only consider them soliton gases with a certain probability. This probability is lower for the cases 4 to 6 and 10 to 11, where the phase information has a larger impact on the soliton energy ratios, leading to wide distributions. For the cases 7 to 8, the phases appear to be less important as the distributions are narrower.

**Table 2**
Parameters related to the narrow bandwidth, weak nonlinearity and deep water assumptions of the NLS equation for the three soliton gases at Eluanbi station. $\Delta f$ denotes the 3 dB bandwidth, $f_{\max}$ the peak frequency and $H_s$ the significant wave height. $k$ was determined from $f_{\max}$ using the deep water dispersion relation. Solitons have been observed for $\Delta f/f_{\max} \lesssim 0.25$ and $kH_s \lesssim 0.3$ in the literature [66]. The value of $kh$ should be at least 1.31.

| Case | $\Delta f/f_{\max}$ | $kH_s$ | $kh$ |
|---|---|---|---|
| 1 | 0.17 | 0.24 | 14.65 |
| 2 | 0.23 | 0.19 | 13.54 |
| 3 | 0.25 | 0.26 | 11.74 |

## Conclusion and outlook

Summarizing, we have analyzed field measurements from buoys in Taiwan waters and identified 11 sea states with extremely high soliton energy ratios. These sea states are associated with relatively small wave heights and short peak periods. Their highly nonlinear character is evidenced by large wave steepness and elevated Benjamin–Feir Index (BFI) values, while skewness and kurtosis do not differ significantly from other conditions. These properties are consistent with highly nonlinear NLS soliton seas previously observed in Currituck Sound, North Carolina [12]. To investigate the influence of directional interference on the soliton energy ratios, we applied a directional filtering technique, which reconstructs wave fields from measured directional power spectra using random phases and wave numbers obtained from the deep water dispersion relation. For each measurement, we simulated wave fields with 100 random phase realizations to assess the distribution of the soliton energy ratio under directional truncation.

*The soliton energy ratios consistently remain high for all three cases measured at Eluanbi station even after directional interference has been removed, which qualifies them as soliton gases in our framework.* We believe that this is the first time that soliton gases have been identified in the open ocean. The cases measured at the Gueishandao and Xiaoliuqiu

stations may or may not qualify as soliton gases with directionality taken into account. Due to the probabilistic nature of our directional filtering technique and the wideness of the distributions obtained for the cases, no definite answer can be given for these cases.

The reconstruction of the wave fields utilized directional power spectra derived from single-point buoy measurements. It was not exact due to the use of random phases and the deep water dispersion relation. Future work should therefore extend the present analysis to other measurement systems, such as radar or stereo-video systems, to obtain spatial data that can better resolve directional structure and phase information. Such data could also be used to verify the kinetic equations for soliton gases. Another important direction is to examine the influence of wind input and wave dissipation, as well as other factors that break the integrability assumption behind the NFT.

## CRediT authorship contribution statement

**Yu-Chen Lee:** Writing – review & editing, Writing – original draft, Visualization, Validation, Software, Resources, Methodology, Investigation, Funding acquisition, Formal analysis, Data curation, Conceptualization. **Sander Wahls:** Writing – review & editing, Validation, Supervision, Software, Resources, Project administration, Methodology, Funding acquisition, Formal analysis, Conceptualization.

## Declaration of competing interest

The authors declare that they have no known competing financial interests or personal relationships that could have appeared to influence the work reported in this paper.

## Acknowledgments

The first author gratefully acknowledges the Ministry of Education, Taiwan, for funding his studies at TU Delft through the 2019 Key Fields Scholarship Program.

## Data availability

The data that support the findings of this study are available from the first author upon reasonable request.